\RequirePackage{fix-cm}
\documentclass{svjour3}                     
\smartqed  
\usepackage{amsmath,amssymb}
\usepackage{latexsym}
\usepackage{graphicx}
\usepackage{epstopdf}

\journalname{Quantum Information Processing}
\begin{document}

\title{Multiparty quantum key agreement protocol secure against collusion attacks}
\author{ Zhiwei Sun \and Xiaoqiang Sun \and Ping Wang}

\institute{Zhiwei Sun \at Shenzhen Key Laboratory of Media Security, Shenzhen University, Shenzhen, 518060, P.R.China
\and Zhiwei Sun \and Xiaoqiang Sun \and Ping Wang  \at College of Information Engineering, Shenzhen University, Shenzhen, Guangdong 518060, P.R.China \\
\email{sunzhiwei1986@gmail.com, wangping@szu.eud.cn}
}

\date{Received: date / Accepted: date}

\maketitle

\begin{abstract}
The fairness of a secure multi-party quantum key agreement (MQKA) protocol requires that all involved parties are entirely peer entities and can equally influence the outcome of the protocol to establish a shared key wherein no one can decide the shared key alone.
However, it is found that parts of the existing MQKA protocols are sensitive to collusion attacks, i.e., some of the dishonest participants can collaborate to predetermine the final key without being detected.
In this paper, a multi-party QKA protocol resisting collusion attacks is proposed. Different from previous QKA protocol resisting $N-1$ coconspirators or resisting $1$ coconspirators, we investigate the general circle-type MQKA protocol which can be secure against $t$ dishonest participants' cooperation. Here, $t < N$.
We hope the results of the presented paper will be helpful for further research on fair MQKA protocols.

\end{abstract}

\section{Introduction}
\label{intro}
Key distribution (KD) allows two authorized participants to establish a shared secret key over a public channel. The shared key can be used for secure communication or authentication protocols.
Key agreement (KA) is another important way to establish keys. Compared with the key distribution, in which one party distributes a secret key to the other, all involved parties in a key agreement protocol can equally influence the outcome of the protocol, and no one or a subset of the group can decide the shared key alone.
One main difference between key agreement and key distribution is that, key agreement protocols not only need to resist adversaries from the outside world, but also are required to prevent the participant attacks.

The security of the classical key agreement protocols are mainly based on the Deffie-Hellman problem or discrete logarithm problem. With the development of quantum computers and the polynomial-time quantum algorithms for prime factorization and discrete logarithm \cite{Shor1994}, the security of classical key agreement protocols have become increasingly vulnerable.

Quantum cryptography, which is based on the quantum mechanical, provides another way for secure key distribution. Since it
can provide unconditional security. It has been developed quickly and become a hot topic in cryptography, such as quantum secret sharing\cite{Hillery1999,Du2012}, quantum secure direct communication\cite{Sun20122,Sun20121}, quantum private comparison\cite{Liuw2014same,Liuw2014improv,Liuw2014CTP,Sun20131,Sun2015} and quantum oblivious transfer \cite{Sun2015PRA}.
Quantum key agreement (QKA) is a new branch of quantum cryptography. Since it was first proposed by Zhou et al. in 2004 \cite{Zhou2004}, lots of QKA protocols have been proposed. In the previously works, only two participants were involved in the QKA protocols \cite{Tsai2009,Chong2011,Chong2010,Huang20141,Shen2014,He2015}. Recently, an enhanced interest on multi-party QKA protocols has been observed \cite{Shi2013,Liu2013,Sun20132,Yin2013,Yin2013three,Chitra2014,Zhu2015,Sun2015cs,Sun2015six,Huang2014}.

Fairness is an important standard needed to be considered in a secure quantum key agreement protocol.
However, it found that most of the quantum key agreement protocols cannot resist collusive attacks \cite{Liu2016}, i.e., parts of the participants of the group can predetermine the shared key before the end of the protocol.
Thus, how to construct a fair and secure key agreement protocol has obtained much attentions.

In this paper, we propose a multiparty quantum key agreement protocol which can resist general collusion attacks. The proposed protocol is based on the idea of our previous multi-party QKA protocol \cite{Sun2015six}.
And the main contribution of the paper is that we present a general way to construct a secure multi-party QKA which can resist $t$ coconspirators. We hope the results of the presented paper will be helpful for further research on fair MQKA protocols.

The rest of this paper is organized as follows. Sect.\ref{sect2} first introduces the formalization of the circle-type multi-party QKA (CT-MQKA) protocols, and the collusion attacks on the CT-MQKA protocols\cite{Liu2016}. Then, we present the MQKA protocol against collusion attacks. Precisely speaking, the presented protocol, which is the generalization of the MQKA protocol in Ref. \cite{Sun2015six}, can resist $t$ coconspirators. Here, $t < N$, where $N$ is the number of the participants in the MQKA protocol. The security and efficiency analyses are given in Sect.3. Sect.4 gives a short conclusion.

\section{Multi-party quantum key agreement protocol}\label{sect2}

We first introduce the formalization of the circle-type multiparty quantum key agreement (CT-MQKA) protocols, and the collusion attacks to the CT-MQKA proposed by Liu \cite{Liu2016}. Then, we show that the CT-MQKA protocols can be used as sub-protocol to construct secure multiparty QKA against collusion attacks. Usually, suppose there are $N$ participants $P_{0}, \cdots, P_{N-1}$, and they have secret bit strings keys $K_{0}, \cdots, K_{N-1}$, respectively.
We denote "$\boxplus$" as addition modular $N$, and "$\boxminus$" as subtraction modular $N$, just like the Ref. \cite{Liu2016} does.

\subsection{Brief review of the CT-MQKA protocol}
At the beginning of the protocol, $P_{i}$ prepares a sequence of entangled states and divides each entangled states into two parts, one of which will be kept, "the home qubit sequence", and the other will be sent out, "the travel qubit sequence".
And, we denote the home qubit sequence as $R_{i}$, and travel qubit sequence as $S_{i}$, respectively, where $i = 0, 1, \cdots, N-1$.

Then all the $S_{i}$s are transmitted in the same direction in the circle.
When all the participants $P_{i\boxplus1}$ have received $S_{i}$, they do the detection and encode their secret keys in the received sequences. Afterwards, they continue to send the above sequence to the next participants. One by one, all the participants will continue the above process. When each travel qubit sequences is sent back to the participant who generated it, i.e., the travel qubit sequence finishes a complete circle, $P_{i}$ can measure $R_{i}$ and $S_{i}$ to get the bitwise exclusive OR results of all the other participant' secret keys. Finally, they can  calculate the final key $K_{final} = \bigotimes_{i=0}^{i=N-1}K_{i}$.

For the convenience of description, we briefly describe the CT-MQKA protocols of Ref.\cite{Sun20132}, which is secure against single participant's attack.
In Ref.\cite{Sun20132}, the whole process of the CT-MQKA is divided into $N$ periods.

In the first period, each $P_{i}$ prepares $R_{i}$ and $S_{i}$ \footnote{For the single state, it can be considered as the entangled states where parts of them $R_{i}$ have already been measured.}, and sends $S_{i}$ to $P_{i\boxplus1}$.
When each $P_{i\boxplus (k-1)}$ receives $S_{i}$, the $k$-th period starts. In the $k$-th period, each $P_{i\boxplus (k-1)}$ performs the detection processes with $P_{i \boxplus (k-2)}$ to detect the possible attacks on $S_{i}$. Then, each $P_{i\boxplus (k-1)}$ encodes his/her secret key $K_{i \boxplus (k-1)}$ on $S_{i}$, and inserts some decoy states in it and sends it to $P_{i\boxplus k}$. When the $k$-th period ends, the $k+1$ period starts. Here, $k= 2, 3, \cdots, N-1$.

In the $N$-th period (a complete circle is finished), each $P_{i}$ performs the attacks detection with $P_{i \boxminus 1}$ as before. After that, the bitwise exclusive OR result of the others' secret keys can be obtained by measuring $R_{i}$ and $S_{i}$.
$P_{i}$  performs the bitwise exclusive OR operation between the above result and $K_{i}$ to get the final key $K_{final}$.

\subsection{Liu's collusive attacks against CT-MQKA protocol}

Liu's collusive attacks can be divided into two stages: the key stealing stage and the key flipping stage \cite{Liu2016}. In the key stealing stage, the collusive participants try to get the bitwise exclusive OR result of the others' secret key in some novel way. Then, they try to flip the encoded secret keys according to the above result to control the final key in the key flipping stage.

And, Ref.\cite{Liu2016} shows that any two participants $P_{n}$ and $P_{m}$ $(n >m)$ are enough to totally control the final key, as long as their position in the circle satisfy the following conditions:

\begin{equation}
n-m = \frac{N}{2} \text{\qquad for an even N};
\end{equation}
\begin{equation}
n-m = \frac{N-1}{2} \text{or} \frac{N+1}{2} \text{\qquad for an odd N}.
\end{equation}

When the above conditions are satisfied, $P_{n}$ and $P_{m}$ perform the following collusion attacks:
\begin{enumerate}
  \item \text{\bf The key stealing stage:}
  \begin{itemize}
    \item In the first period, $P_{n}$ and $P_{m}$ share all the information about $R_{n}$, $S_{n}$, $K_{n}$ and $R_{m}$, $S_{m}$, $K_{m}$ and the value of the expected key $K_{expected}$.
    \item In the $(n-m)$-th period which started by $P_{m}$, $P_{n}$ has received the travel sequence $S_{m}$. Combined with the shared information about $R_{m}$, $S_{m}$, $P_{n}$ can obtain the bitwise exclusive OR result of the secret key $K_{m+1}, K_{m+2}, \cdots, K_{n-1}$ by measuring $R_{m}$ and $S_{m}$. Similarly, $P_{m}$ can get the bitwise exclusive OR result of the secret key $K_{n+1}, K_{n+2}, \cdots, K_{m-1}$ by measuring $R_{n}$ and $S_{n}$ in the $(N-n+m)$-th period which started by $P_{n}$.
    \item $P_{n}$ ($P_{m}$) sends the above bitwise exclusive OR result to $P_{m}$ ($P_{n}$) immediately he/she gets it.
  \end{itemize}
  \item \text{\bf The key flipping stage:} In the $\frac{N}{2}$ period (for the convenience of description of the collusion attacks, suppose $N$ is an even number), each of $P_{n}$ and $P_{m}$ gets the bitwise exclusive OR result of half of the others' secret key. After exchanging with each other, they get the legal final key $K_{final}$ ahead of others.
      Then $P_{n}$ and $P_{m}$ can predetermine the final key by encode $K_{n}^{'} = K_{n} + K_{expected}+K_{final}$ instead of $K_{n}$, and $K_{m}^{'} = K_{m} + K_{expected}+K_{final}$ instead of $K_{m}$ respectively in the rest periods. It can be verified that, in the last period, for any participant $P_{i}$, he/she will get the final key is $K_{final}^{'}=K_{expected}$.
\end{enumerate}

\subsection{Multi-party QKA protocol against $t$ coconspirators}
Recently, Sun et. al. proposed a novel multi-party quantum key agreement protocol by using entangled states \cite{Sun2015six}, which is secure against $2$ collusion attackers. In their protocol, each participant sends out two sequences, instead of one sequence in CT-MQKA. Each of the two sequences "runs" half circle. Two collusive participants cannot succeed any more by using Liu's collusion attacks. Because each of them can only get the bitwise exclusive OR result of half of the other's personal keys after the last period, which leaves no time for them to flip the others' sequences.
However, when three participants collaborate with each other, they can be succeed. Thus, Sun et. al.'s protocol can be only secure against $2$ coconspirators.
Even though more than two participants can succeed in attacking Sun et. al.'s MQKA protocol, it provides a new perspective for in-depth analysis of multi-party QKA protocols secure against collusion attacks.

Suppose there are $N$ participants involved in the multi-party QKA protocol. And, we hope it can resist $t$ dishonest participants' cooperation, where $t \leq \frac{N}{2}$. And the $N$ participants are arranged uniformly in the circle.
The proposed multi-party QKA protocol against $t$ coconspirators is described as follows.

\begin{figure}[h]
\centering
\includegraphics[width=9cm,height=8cm]{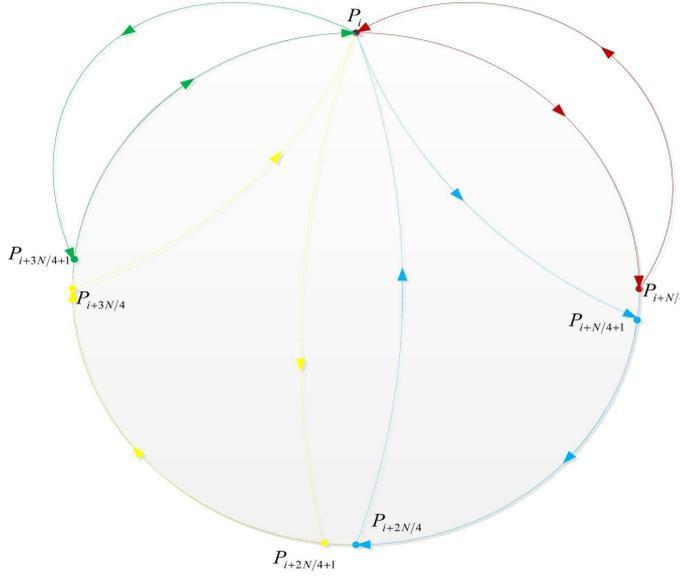}
\caption{We give an example for $t=4$, i.e., the MQKA can resist $4$ collusive attackers. The circle is divided into $4$ parts, the red part, the blue part, the yellow part and the green part. Each part is a complete sub-circle when $S_{i}$ is returned back to $P_{i}$.}
\label{fig1}
\end{figure}

\begin{enumerate}
  \item In the first period, each $P_{i}$ prepares $t$ sequences of entangled states, and divides each entangled states into two parts $(R_{i}^{0}, S_{i}^{0})$, $\cdots$, $(R_{i}^{t-1}, S_{i}^{t-1})$ \footnote{For the single state, it can be considered as the entangled states where parts of them have already been measured.}, respectively, and sends $S_{i}^{0}$ to $P_{i \boxplus 1}$, $S_{i}^{1}$ to $P_{i\boxplus\lfloor \frac{N}{t}\rfloor \boxplus 1}$, $\cdots$, $S_{i}^{t-1}$ to $P_{i\boxplus\lfloor \frac{(t-1)N}{t}\rfloor \boxplus 1}$. Here, $\lfloor x\rfloor$ represents the maximum integer which is not more than $x$. For the convenience of description, we simply write $\frac{N}{t}$ instead of $\lfloor \frac{N}{t}\rfloor$. Note that $P_{i}$ divides the circle into $t$ parts, each part has $\frac{N}{t}$ participants.
 \item\label{2} \textbf{Detection phase} When each $P_{i\boxplus (k-1)}$ receives $S_{i}^{0}$, $P_{i\boxplus \frac{N}{t} \boxplus (k-1)}$ receives $S_{i}^{1}$, $\cdots$, $P_{i\boxplus \frac{(t-1)N}{t} \boxplus (k-1)}$ receives $S_{i}^{t-1}$, respectively, the $k$-th period starts. Here, $k= 2$. \\
     In the $k$-th period, each $P_{i\boxplus (k-1)}$ first performs the detection processes with $P_{i \boxplus (k-2)}$ to detect the possible attacks on $S_{i}^{0}$, $P_{i\boxplus \frac{N}{t} \boxplus (k-1)}$ first performs the detection processes with $P_{i \boxplus \frac{N}{t} \boxplus (k-2)}$ to detect the possible attacks on $S_{i}^{1}$, $\cdots$, $P_{i\boxplus \frac{(t-1)N}{t} \boxplus (k-1)}$ first performs the detection processes with $P_{i \boxplus \frac{(t-1)N}{t} \boxplus (k-2)}$ to detect the possible attacks on $S_{i}^{t-1}$, respectively.\\
     Note that, in the second period, $P_{i\boxplus 1}$, $P_{i\boxplus \frac{N}{t} \boxplus 1}$, $\cdots$, $P_{i\boxplus \frac{(t-1)N}{t} \boxplus 1}$ perform detection process with $P_{i}$, instead of their former participant in the circle, to detect the possible attacks.
\item\label{3} \textbf{Encoding Phase} When all the sequences are secure, each $P_{i\boxplus (k-1)}$ encodes his/her secret key $K_{i \boxplus (k-1)}$ in $S_{i}^{0}$, and inserts some decoy states in it and sends it to $P_{i\boxplus k}$, $P_{i\boxplus \frac{N}{t} \boxplus (k-1)}$ encodes his/her secret key $K_{i\boxplus \frac{N}{t} \boxplus (k-1)}$ in $S_{i}^{1}$, and inserts some decoy states in it and sends it to $P_{i\boxplus \frac{N}{t} \boxplus k}$, $\cdots$, $P_{i\boxplus \frac{(t-1)N}{t} \boxplus (k-1)}$ encodes his/her secret key $K_{i\boxplus \frac{(t-1)N}{t} \boxplus (k-1)}$ in $S_{i}^{t-1}$, and inserts some decoy states in it and sends it to $P_{i\boxplus \frac{(t-1)N}{t} \boxplus k}$, respectively.
\item The parties sequentially execute eavesdropping check and the encoding processes in the same way as participants did in steps \ref{2} and \ref{3}. When the $k$-th period ends, the $k+1$ period starts. Here, $k= 2, \cdots, \frac{N}{t}$.
\item In the $\frac{N}{t} + 1$-th period (a complete sub-circle is finished, for example Fig.\ref{fig1}), each $P_{i\boxplus \frac{N}{t}}$, $P_{i\boxplus \frac{2N}{t}}$, $\cdots$, $P_{i\boxplus t \lfloor \frac{N}{t}\rfloor }$ performs the attacks detection with $P_{i}$, respectively,
    After that, the bitwise exclusive OR result of the others' secret keys can be obtained by measuring $(R_{i}^{0}, S_{i}^{0})$, $\cdots$, $(R_{i}^{t-1}, S_{i}^{t-1})$. $P_{i}$  performs the bitwise exclusive OR operation between the above result and $K_{i}$ to get the final key $K_{final}$.
\end{enumerate}

\section{Security and Efficiency analysis}
In this section, we will give the security and efficiency analysis of the proposed multi-party QKA protocol.

\subsection{Security analysis}
We first consider $t> \frac{N}{2}$.
When $N > t > \frac{N}{2}$, we have $1 < \frac{N}{t} <2$, i.e., $\lfloor \frac{N}{t} \rfloor =1$. In this case, the circle will be divided into $N$ parts, each part has $1$ participant.
Then, this kind of CT-MQKA protocol becomes the complete-graph-type MQKA (CGT-MQKA) protocol \cite{He2015,Liu2013,Yin2013three}, which has been proven fair against both single and collusion attacks.

When $t < \frac{N}{2}$.
For the simplest case $t=1$, the proposed multi-party QKA protocol becomes the standard circle-type multi-party QKA (CT-MQKA) protocol \cite{Sun20132,Yin2013,Sun2015cs}, which has been proven that it is secure against single participant attack.

For the general case, the security analysis is similar to the security analysis of the MQKA protocol resisting $2$ coconspirators \cite{Sun2015six}.
As we know, Liu's collusive attacks can be divided into two stages: the key stealing stage and the key flipping stage. In the key stealing stage, the collusive participants try to get the bitwise exclusive OR result of the others' secret key. Then, they can flip the encoded secret keys according to the above result to control the final key in the key flipping stage.
In order to resist Liu's collusion attacks, the key stealing stage or the key flipping stage must be destroyed.
It can be verified that the proposed protocol cannot resist collusion attack at the key stealing stage. In other words, $t$ participants, in the special positions of the circle, can get the final key ahead of others, by using Liu's collusion attacks.
However, when the key stealing stage is finished, the whole protocol is also accomplished, i.e., the collusive participants have no time to flip the encoded secret keys. The key flipping stage is destroyed.
Thus, the $t$ coconspirators cannot predetermine the final key any more. For the precise security analysis, it can refer to Ref. \cite{Liu2013,Sun2015six}.

\subsection{Efficiency analysis}
We use the qubit efficiency to measure the efficiency of the proposed MQKA protocol. The qubit efficiency was introduced by Cabello \cite{Cabello2000} in 2000, which is given as
\begin{eqnarray}
 \eta =\frac{c}{q+b},
\end{eqnarray}
where $c$ denotes the length of the transmitted message bits (the length of the final key), $q$ is the number of the used qubits, and $b$ is the number of classical bits exchanged for decoding of the message (classical communication used for checking of eavesdropping is not counted).

In order to generate $n$ bits of shared key, each party has to prepare $t.n$ single photons and $\kappa t. n$
decoy particles in the proposed protocol. There is no classical bits exchanged for decoding of the shared key.
Hence, the qubit efficiency of proposed protocol can be computed, $\eta = \frac{n}{(tn +\kappa tn)N}=\frac{1}{(\kappa+1)tN}$, where $\kappa$ is the detection rate and $N$ is the number of the participants.
It can be verified that when $t=N-1$, the qubit efficiency is  $\frac{1}{(\kappa+1)N(N-1)}$, which is identical to the qubit efficiency of Ref. \cite{Liu2013}. This also implies that the proposed protocol is a general case of MQKA protocol resisting $t$ coconspirators.

\section{Conclusion}

In conclusion, we propose a multiparty quantum key agreement protocol which can resist collusion attacks which is presented in the Ref. \cite{Liu2016}. The proposed protocol is based on the idea of our multi-party QKA protocol which can resist 2 coconspirators \cite{Sun2015six}. And the main contribution of the paper is that we present a general way to construct a secure multi-party QKA which can resist $t$ participants collaborating to predetermine the final key, which protects the honest participants' fairness. We hope the results of the presented paper will be helpful for further research on more secure and more fair MQKA protocols.

\begin{acknowledgements}
This work is funded by the National Natural Science Foundation of China (No. 61402293, 61300204), the Science and Technology Innovation Projects of Shenzhen (No. JCYJ20150324141711665 and No. JCYJ20150324141711694), Natural Science Foundation of SZU (No. 201435), Shenzhen R$\&$D Program (GJHZ20140418191518323), Seed Funding from Scientific and Technical Innovation Council of Shenzhen Government (No. 827-000035), Natural Science Foundation of Guangdong (2015A030313630), Opening Project of Shanghai Key Laboratory of Integrated Administration Technologies for Information Security, and China Postdoctoral Science Foundation (No. 2015M572360).
\end{acknowledgements}



\end{document}